\documentclass[useAMS,usenatbib]{mn2e}
\usepackage{graphicx}
\usepackage{booktabs}
\usepackage{subfigure}
\usepackage{ulem}
\usepackage{amsmath,amssymb,latexsym}

\def \aap {A\&A}

\def \aj {AJ}
\def \an {AN}
\def \apj {ApJ}
\def \mnras {MNRAS}

\def \kmps {\mathrm{km}\,\mathrm{s}^{-1}}
\def \vsini {v\sin i}

\def \Msun {\mathrm{M}_\odot}

\def \Lsun {\mathrm{L}_\odot}
\def \Rsun {\mathrm{R}_\odot}
\def \RV {\mathrm{RV}}
\def\softL{L\kern-0.8ex\raise0.1ex\hbox{'}\kern0.1ex}

\title[V501 Aur]{On the nature of the candidate T-Tauri star V501\,Aurigae
\thanks{Based on observations obtained with telescopes of the University Observatory Jena, operated 
by the Astrophysical Institute of the Friedrich-Schiller-University Jena, 
at Michael Adrian Observatory, Germany and at Star\'a Lesn\'a and Kolonica Observatory, Slovakia. 
Based on the data from SuperWASP and NSVS archives.}}

\author[M.\,Va\v{n}ko et al.]
{M.\,Va\v{n}ko,$^1$\thanks{vanko@ta3.sk}
G.\,Torres$^2$, {\softL}.\,Hamb\'alek$^1$, T.\,Pribulla$^1$, L.A.\,Buchhave$^3$, J.\,Budaj$^1$,
\newauthor
P.\,Dubovsk\'y$^4$, Z.\,Garai$^1$, C.\,Ginski$^5$, K.\,Grankin$^6$, R.\,Kom\v{z}\'{\i}k$^1$,
V.\,Krushevska$^7$, 
\newauthor
E.\,Kundra$^1$, C.\,Marka$^8$, M.\,Mugrauer$^9$, R.\,Neuh\"auser$^9$, J.\,Ohlert$^{10,11}$,
\v{S}.\,Parimucha$^{12}$, 
\newauthor
V.\,Perdelwitz$^{13}$, St.\,Raetz$^{14}$, S.Yu.\,Shugarov$^{1,15}$\\
 $^1$Astronomical Institute, Slovak Academy of Sciences, 059 60 Tatransk\'a Lomnica, Slovakia\\
 $^2$Harvard-Smithsonian Center for Astrophysics, 60 Garden St., Cambridge, MA 02138, USA\\
 $^3$Centre for Star and Planet Formation Natural History Museum of Denmark, University of Copenhagen, DK-1350 Copenhagen, Denmark\\
 $^4$Vihorlat Observatory, Mierova 4, Humenn\'e, Slovakia\\
 $^5$Leiden Observatory, Leiden University, P.O. Box 9513, 2300 RA Leiden, The Netherlands\\
 $^6$Crimean Astrophysical Observatory, Scientific Research Institute, 298409, Nauchny, Crimea\\
 $^7$Main Astronomical Observatory, National Academy of Sciences of Ukraine, 27, Akademika Zabolotnoho, Kyiv, 03680, Ukraine\\
 $^8$Instituto de Radioastronom\'ia Milim\'etrica, Av. Divina Pastora 7, N\'ucleo Central, E-18012 Granada, Spain\\
 $^9$Astrophysikalisches Institut und Universit\"ats-Sternwarte, Schillerg\"a{\ss}chen 2-3, 07745 Jena, Germany\\
 $^{10}$University of Applied Sciences, Wilhelm-Leuschner-Strasse 13, 61169 Friedberg, Germany\\
 $^{11}$Michael Adrian Observatory, Astronomie Stiftung Trebur, Fichtenstrasse 7, 65468 Trebur, Germany\\
 $^{12}$Institute of Physics, Faculty of Science, University of P.J. \v{S}af\'arik in Ko\v{s}ice, Park Angelinum 9, 04001 Ko\v{s}ice, Slovakia\\
 $^{13}$Hamburger Sternwarte, Gojenbergsweg 112, 21029 Hamburg, Germany\\ 
 $^{14}$Freiburg Institute of Advanced Studies (FRIAS), University of Freiburg, Albertstra{\ss}e 19, D-79104 Freiburg, Germany\\
 $^{15}$Sternberg Astronomical Institute, Moscow State University, Universitetskij pr., 13, Moscow, 119991, Russia\\}

\begin{document}

\date{Accepted XXXX December 15. Received 2016 August 23; in original form
2016 August 23}

\pagerange{\pageref{firstpage}--\pageref{lastpage}} \pubyear{2017}

\maketitle

\label{firstpage}

\begin{abstract}
We report new multi-colour photometry and high-resolution
spectroscopic observations of the long-period variable V501\,Aur,
previously considered to be a weak-lined T-Tauri star belonging to the
Taurus-Auriga star-forming region. The spectroscopic observations
reveal that V501\,Aur is a single-lined spectroscopic binary system
with a 68.8-day orbital period, a slightly eccentric orbit ($e \sim
0.03$), and a systemic velocity discrepant from the mean of
Taurus-Auriga. The photometry shows quasi-periodic variations on a
different, $\sim$55-day timescale that we attribute to rotational
modulation by spots. No eclipses are seen. The visible object is a
rapidly rotating ($v \sin i \approx 25~\kmps$) early K star, which
along with the rotation period implies it must be large ($R >
26.3~\Rsun$), as suggested also by spectroscopic estimates indicating
a low surface gravity. The parallax from the {\it Gaia\/} mission and
other independent estimates imply a distance much greater than the
Taurus-Auriga region, consistent with the giant interpretation. Taken
together, this evidence together with a re-evaluation of the
Li\,I~$\lambda$6707 and H$\alpha$ lines shows that V501\,Aur is not a
T-Tauri star, but is instead a field binary with a giant primary far
behind the Taurus-Auriga star-forming region. The large mass function
from the spectroscopic orbit and a comparison with stellar evolution
models suggest the secondary may be an early-type main-sequence star.
\end{abstract}

\begin{keywords}
stars: individual: V501\,Aur.
\end{keywords}

\section{Introduction}
\label{intro}

V501\,Aurigae (W72, 1RXS J045705.7+314234, HD\,282600, TYC~2388-857-1,
$V=10.57$, $B-V=1.62$) was detected as an X-ray source by ROSAT \citep{wichmann96}, and was classified by these authors as a possible
new weak-lined T-Tauri star (hereafter WTTS) based on the presence of
the Li\,I~$\lambda$6707 resonance line in low-resolution optical
spectra, the H$\alpha$ line slightly in emission, and the late
spectral type (K2). \citet{frink97} examined the proper motion of the
star and concluded that it is a likely member of the central region of
the Taurus-Auriga star-forming region (SFR). Additional Li
observations based on higher-resolution spectra of 35 of the candidate
WTTS by \citet{wichmann96} were published by \cite{martin99}, who
reported for V501\,Aur only an upper limit to the Li equivalent width
(EQW) of 90~m\AA, along with an H$\alpha$ EQW of 0.9~\AA\ {\it in
absorption}, and the same spectral classification as the previous
authors.

\citet{wichmann00} revisited many of the candidate WTTS from their
earlier paper including V501\,Aur, drawing on new high-resolution
spectroscopic observations from the Harvard-Smithsonian Center for
Astrophysics (CfA) as well as the ELODIE spectrograph at the
Haute-Provence Observatory. They reported a new Li measurement of
${\rm EQW}_{\rm Li\,I} = 138$~m\AA\ from their ELODIE observations,
and obtained effective temperature estimates of $T_{\rm eff} = 5350$~K
and $T_{\rm eff} = 4897$~K from a cross-correlation analysis of the
CfA spectra and directly from the K2 spectral type, respectively. They
reported also estimates of the rotational velocity as
$\vsini=25~\kmps$ (CfA) and $\vsini=27~\kmps$ (ELODIE). Their Li
measurement was more typical of Pleiades-age stars than younger WTTS,
which, together with the discovery that V501\,Aur is a single-lined
spectroscopic binary as revealed by the CfA spectra, led them to be
more cautious in claiming WTTS status for the object. Nevertheless, on
the tentative assumption that it is a member of Taurus-Auriga and is
therefore at a distance of $\sim$140~pc, they used evolutionary models
by \citet{dantona94} to estimate the luminosity ($L = 7.8~\Lsun$),
radius ($R = 3.86~\Rsun$), and mass ($M = 1.36~\Msun$) of the star,
and assigned it a very young age of $\log({\rm age}) = 5.53$ ($\sim 3
\times 10^5$~yr). Additional H$\alpha$ and projected rotational
velocity measurements were reported by \citet{nguyen09} as ${\rm
  EQW}_\mathrm{{H}\alpha}=1.0\pm0.5$ \AA\ (absorption) and $\vsini =
25.5\pm 1.5~\kmps$, in good agreement with those of \citet{martin99}
and \citet{wichmann00} mentioned above.

Photometric monitoring of V501\,Aur has been carried out by several
authors. \citet{bouvier97} observed it as part of a sample of 58 WTTS
detected in the ROSAT All-Sky Survey (RASS). They were able to derive
rotation periods for 18 of their stars, all but one being ascribed to
rotational modulation by stellar spots. The one exception was
V501\,Aur, which showed evidence of variability on a very long
timescale ($P > 37.6$ days) uncharacteristic of WTTS, and displayed no
appreciable modulation in the $B-V$ colour in their observations, as
would be expected in the spot scenario. They also claimed the star to
be a double-lined spectroscopic binary, though this was based on a
high-resolution but very low signal-to-noise ratio (SNR) spectrum
taken at the Haute-Provence Observatory. \citet{grankin08} presented
a homogeneous set of photometric measurements for WTTS extending up to
20 years. Their data were collected within the framework of the ROTOR
program (Research Of Traces Of Rotation), aimed at the study of the
photometric variability of pre-main-sequence (PMS) objects. The data
set contains rotation periods for 35 out of 48 stars, including
V501\,Aur. Our target was observed in several seasons from 1994 to
2004 (see Section~\ref{photometry}).  The photometry showed wave-like
variability of the object with an average period of about 55 days.

V501\,Aur has also been included in an 8.4~GHz VLA survey of
lithium-rich late-type stars from the RASS by \citet{carkner97}. The
object was detected as a radio source with a radio emission strength
of $S_{\rm 8.4}=0.17\pm0.05$~mJy. \citet{daemgen15} recently used the
NIRI instrument on the 8m Gemini North telescope to carry out a
near-infrared high angular resolution survey for stellar and
sub-stellar companions in the Taurus-Auriga SFR, but reported no
detections around V501\,Aur. Finally, in a brief study by a subset of
the present authors, \citet{vanko15} presented new $VRI$ measurements
confirming the $\sim$55 day photometric periodicity. The photometric data 
obtained at University Observatory Jena and Star\'a Lesn\'a Observatory between 
2007 and 2013 were used. Based on the CfA spectroscopy (1996-1997) the authors found
that V501\,Aur is a single-lined spectroscopic binary with a
nearly circular orbit, a large mass function implying a fairly massive
companion, and an orbital period of 68.8 days that is distinctly longer than  
the photometric period. They speculated that the unseen
companion may be a binary, or alternatively that the primary star may
be a giant (implying a much greater distance than previously assumed),
which might also explain the lack of detection of a main-sequence
secondary.

Here we present additional spectroscopic and photometric observations
of V501\,Aur that motivate us to revisit the object with the following
goals: (i) to improve the determination of its orbital elements as
well as its physical parameters, including the atmospheric properties
(temperature, surface gravity, and the strength of the Li and
H$\alpha$ lines); (ii) to better characterize the photometric
variability, which is unusual for a WTTS, through a comprehensive
study of all available observations; (iii) to investigate the
difference between the photometric and orbital periods and its
implications; and (iv) to present a coherent picture of the true
nature of the system based on all available information, including a
recent estimate of the parallax of V501\,Aur from {\it Gaia} that
appears to conflict with the notion of membership in the Taurus-Auriga
star-forming region.
       	
The paper is organized as follows. In Section~\ref{photometry} we
present our new photometric observations followed by a detailed period
analysis. Section~\ref{spectroscopy} describes our new spectroscopic
observations and reports an updated spectroscopic orbital solution.
Section~\ref{extinction} contains a discussion of interstellar
reddening. In Section~\ref{properties} we review the physical
properties of V501\,Aur and re-examine the evidence for membership in
the Taurus-Auriga SFR, presenting a coherent picture of its
evolutionary state based on stellar evolution models. We conclude in
Section~\ref{conclusion} with our final thoughts.

\section{Photometric observations}
\label{photometry}

The differential photometry used in this paper was carried out between
2007 and 2016 at four different observatories, two in Slovakia and two
in Germany. The two in Slovakia are the Star\'a Lesn\'a Observatory
(G1; 49\degr 09\arcmin 10\arcsec N, 20\degr 17\arcmin 28\arcsec E) and
the Kolonica Observatory (KO; 48\degr 56\arcmin 06\arcsec N, 
22\degr 16\arcmin 26\arcsec E). The two observatories in Germany are the University Observatory
Jena (GSH; 50\degr 55\arcmin 44\arcsec N, 11\degr 29\arcmin 03\arcsec
E) and the Michael Adrian Observatory (MAO; 49\degr 55\arcmin
31\arcsec N, 08\degr 24\arcmin 41\arcsec E).  All of the observations
used Johnson-Cousins ($UBVR_{C}I_{C}$) and Bessel ($UBVRI$) filter
sets.  More detailed information on the individual observatories and
instruments is given in Table~\ref{tab01}.

\begin{table}
\caption{Overview of telescopes and instruments/detectors used to obtain the photometry of V501\,Aur. 
$D/f$ gives the diameter and focal length of the telescope. FoV is the field of view of the instrument. 
Observatory abbreviations: G1 -- pavilion of the Star\'a Lesn\'a Observatory, 
GSH -- University Observatory Jena (see \citealp{mugi09, mugi10}), MAO -- Michael Adrian Observatory in Trebur, 
KO -- Kolonica Observatory \citep[ZIGA telescope; see][]{parimucha15}.
\label{tab01}}
\footnotesize 
\begin{center}
\begin{tabular}{lccc}
\hline
Obs. & Telescope &  Detector & FoV      \\
     &  $D/f$    &  CCD size &          \\
     &  [mm]     &           & [arcmin] \\
\hline
G1          & Newton    &  SBIG ST10-MXE                   & 20.4$\times$13.8 \\
            & 508/2500  &  2184$\times$1472, 6.8 $\mu$m  &                    \\
            & Cassegrain& FLI ML 3041                      & 14$\times$14     \\
            & 600/7500  &  2048$\times$2048, 15 $\mu$m   &                    \\
            & Maksutov  & SBIG ST10 MXE                    & 28.5$\times$18.9 \\
            & 180/1800  & 2184$\times$1472, 6.8 $\mu$m     &                  \\
KO          & Cassegrain&  MI G4-16000                     & 36$\times$36     \\
            & 508/3454  &  4096$\times$4096, 9 $\mu$m    &                    \\
MAO         & Cassegrain&  SBIG STL-6303E                  & 10$\times$7      \\
            & 1200/9600 &   3072$\times$2048, 9 $\mu$m   &                    \\ 
GSH         & Cassegrain&  SITe TK1024                     & 37.7$\times$37.7 \\
            & 250/2250  &  1024$\times$1024, 24 $\mu$m   &                    \\ 
            & Schmidt   &  E2V CCD42-10                    & 52.8$\times$52.8 \\
            & 600/1350  &  2048$\times$2048, 13.5 $\mu$m &                    \\
\noalign{\smallskip}
\hline
\end{tabular}
\end{center} 
\end{table}  

The CCD frames were subjected to standard photometric corrections
(overscan, dark, and flatfield), and we then performed aperture
photometry using tasks within {\tt IRAF\/}\footnote{IRAF is
  distributed by the National Optical Astronomy Observatories, which
  are operated by the Association of Universities for Research in
  Astronomy, Inc., under cooperative agreement with the National
  Science Foundation.} (for G1 and GSH), the {\tt C-munipack\/}
package\footnote{http://c-munipack.sourceforge.net/} (KO), and {\tt
  Mira\_Pro\_7\/}\footnote{http://www.mirametrics.com/mira\_pro.htm}
(MAO).  The comparison star for all observations was HD~282599.

Additional photometric data used for this study were taken from the
SuperWASP\footnote{http://wasp.cerit-sc.cz/form} and NSVS
archives\footnote{http://skydot.lanl.gov/nsvs/nsvs.php}. The WASP
instruments have been described by \citet{pollacco06}, and the
reduction techniques discussed by \citet{smalley11} and
\citet{holdsworth14}. The aperture-extracted photometry from each
camera on each night was corrected for atmospheric extinction,
instrumental colour response, and system zero-point relative to a
network of local secondary standards. The resulting pseudo-$V$
magnitudes are comparable to {\it Tycho-2\/} \citep{hog00} $V$
magnitudes \citep{butters10}. The WASP observations for V501\,Aur are
from the first data release (DR1) of the WASP archive, which contains
light curves from 2004 to 2008.

NSVS data for V501~Aur are available for the years 1999--2000 (object
IDs 6830118 and 6841841). The NSVS magnitudes were converted to the
Johnson $V$ system using
\begin{equation}
 \label{eq:nsvsmag}
  m_V=\frac{1.875\times m_{\rm NSVS}+m_B}{2.875}~,
\end{equation}
where $m_B=12.16$ is the Johnson $B$ magnitude of V501\,Aur \citep{hog00}. Numerical constants were adopted from \citet{woz04}. 

Prior to the analysis the light curves from individual
observatories/instruments were corrected for small magnitude offsets,
with the light curve from WASP being taken as the reference.
The photometric precision of data points from all sources was in the range of 0.001--0.021 mag in the $V$ passband.
The worst precision was achieved in the SuperWASP data.

\subsection{Period analysis}
\label{period}

We have defined an observational season by the observability of our target star beginning in August and ending in April of the 
following year (see Table~\ref{tab:seasons}). The data from the first three seasons are mainly from the SuperWASP archive, 
and the data coverage is much greater than in later seasons. Season ``0" contains only NSVS data.

\begin{table*}
  \begin{center}
  \caption{Observational seasons of V501\,Aur with corresponding periods and amplitudes  
of the wave-like variability (sine model) in the $V$ passband. Periods marked with an asterisk ``*'' were 
obtained from the wavelet analysis.}
  \label{tab:seasons}
  \begin{tabular}{lcccccccl}
  \hline
  Season & Start & End & Duration & Nights & Points & Period & Amplitude & Source \\
         &       &     & [d]      &        &        & [d]    & [mag]     &        \\
  \hline
  0 & 1999-08-24 & 2000-03-26 & 214.8 & 52 &  112 & 58.07(3)  & 0.019  & NSVS  \\
  1 & 2004-08-02 & 2004-09-30 &  59.0 & 43 &  912 & 59.22(4)  & 0.013  & SuperWASP\\
  2 & 2006-09-11 & 2007-02-15 & 157.7 & 63 & 2393 & 55.90(3)  & 0.046  & SuperWASP\\
  3 & 2007-09-29 & 2008-04-23 & 207.7 & 63 & 3165 & 54.92(3)  & 0.019  & SuperWASP/GSH/G1 \\
  4 & 2008-10-03 & 2009-04-06 & 185.7 & 12 &  100 & 55.72(3)* & -      & GSH    \\
  5 & 2010-09-23 & 2011-04-16 & 204.8 & 31 &  602 & 56.34(3)  & 0.040  & G1    \\
  6 & 2011-08-26 & 2012-01-03 & 130.6 &  6 &   84 & 53.96(3)* & -      & G1    \\
  7 & 2015-11-23 & 2016-03-31 & 128.9 & 15 &  719 & 59.60(4)  & 0.065  & G1/KO/MAO \\
  &&&&&&&\\
  $\Sigma$ & 1999-08-24 & 2016-03-31 & 6063.9 & 285 & 8087 & 55.45(3)  & 0.026 & \\
  \hline
  \end{tabular}
  \end{center}
\end{table*}

For the period analysis we used the {\tt
  VStar}\footnote{https://www.aavso.org/vstar-overview} package
developed by the American Association of Variable Star Observers
(AAVSO). The period analysis was performed using a Date Compensated
Discrete Fourier Transform (DC DFT) algorithm \citep{FM81}. This method
compensates for gaps within the data set using weighting,
discriminating aliases, and allowing for frequency harmonic
filtering.
At first, we have applied DC DFT analysis on all SuperWASP data
(because they have the best sampling and coverage) to explore wide range 
(0.00007 to 100 days) of possible periods in the dataset.
The lowest frequency roughly corresponds to $1/(4w)$, where $w$ was the span/window of the data.
The highest frequency was set to the median interval between consecutive datapoints.
The step in frequency was set equal to the lowest frequency.

\begin{figure*}
\centering   
\includegraphics[width=170mm,clip=0]{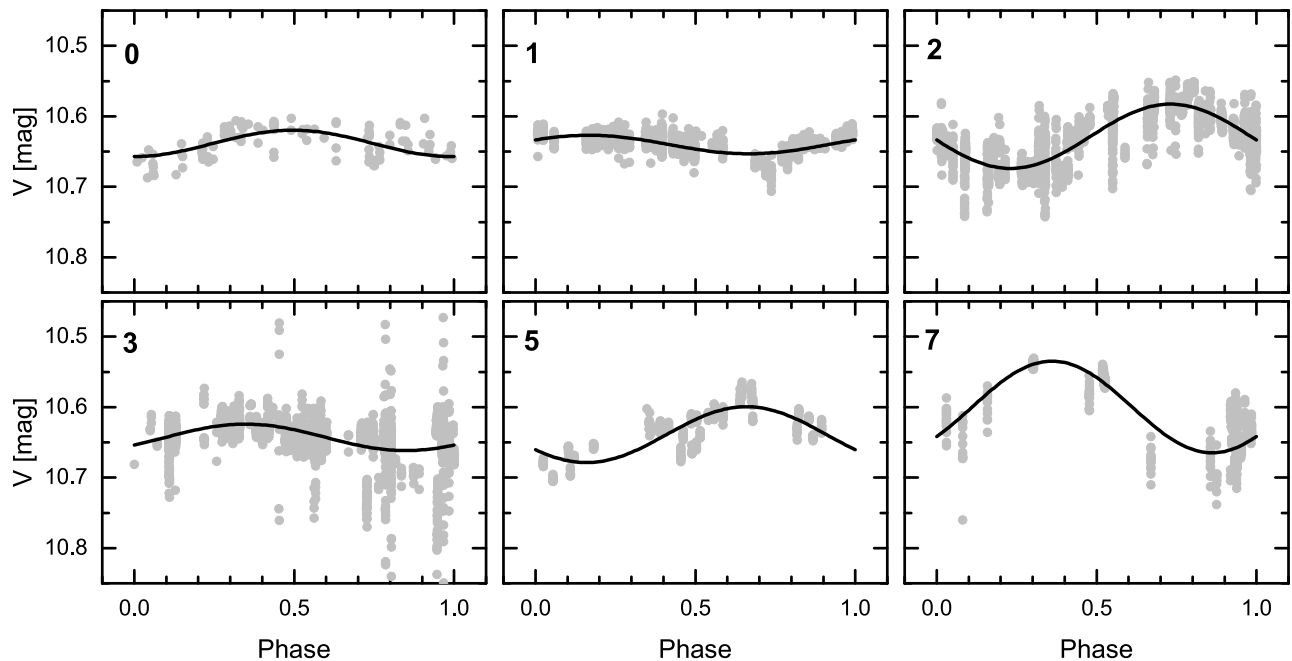}
\caption{Results of our period analysis in individual observing seasons
(indicated by the panel number). No significant period was found in 
Seasons~4 and 6. Each phase-folded light curve comprises of
two to four cycles. The data points are distributed rather uniformly in phase in each cycle.}
\label{fig:periods}
\end{figure*}

No significant periods were found outside 40 -- 70 days range for all observing seasons.
Figure~\ref{fig:periods} presents folded light curves at the periods
found for each season.

\begin{figure}
  \centering  
\includegraphics[width=82mm,clip=4]{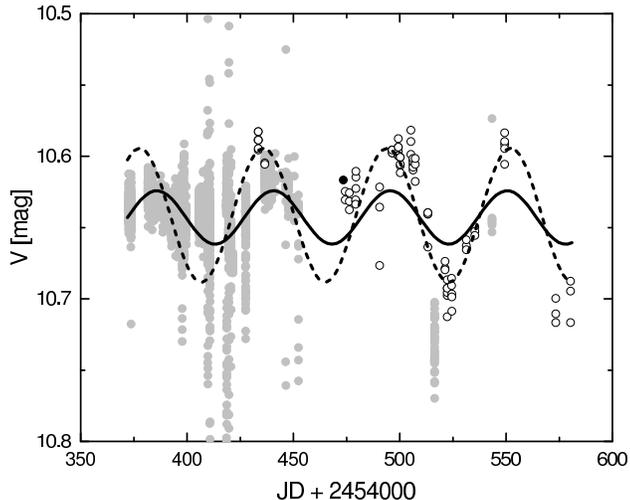}
  \caption{Fit for the period $P_3\sim54.92$\,days (solid line) based on all data
           from the observational Season 3 and fit for the period $P\sim58.2$\,days
           found from data of this work only. The observations by SuperWASP, GSH and
           G1 data are shown as grey, empty and black points, respectively.
           \label{fig:season3}}
\end{figure}

Season~3 (2007--2008) produced interesting results. The data from this season are merged SuperWASP, GSH, and G1 photometry, 
and the coverage is better in the SuperWASP data (see Figure~\ref{fig:season3}). It is important to note that only $\sim2\%$ of
data points are overlapping in this sample.
The period analysis of the entire season resulted in a period of $P_3=54.92$ days. However, our new data from 2008 (no SuperWASP) 
were fitted poorly by this period. No other significant period was found in the full data set (our data $+$ SuperWASP) of Season~3. 
As a test, we discarded all SuperWASP data from this season and ran an independent period analysis on the remaining data 
(our own), which yielded a period $P_3'=58.21$ days. Because the number of SuperWASP data points is $\sim$40 times larger 
than the number of our own observations, we have chosen to retain only period $P_3$ for further study. 
We note, however, that the presence of $P_3'$ is difficult to understand simply by undersampling of the data.

The original table of results by \citet{grankin08} showed that between 1994--1995 and 1996--1997 the observed period of the photometric wave of 
V501\,Aur changed significantly. If a similar change occurred during our Season~3, this could perhaps explain the presence of a second periodicity. 
For Seasons~4 and 6 we were unable to find a significant periodicity because of considerable undersampling (see Table~\ref{tab:seasons}) and large 
gaps in the data. Below in Figure~\ref{fig:wwz-all} we compare the periods determined in this work to those obtained by \citet{grankin08}.

\begin{figure*}
 \centering
 \includegraphics[width=170mm,clip=0]{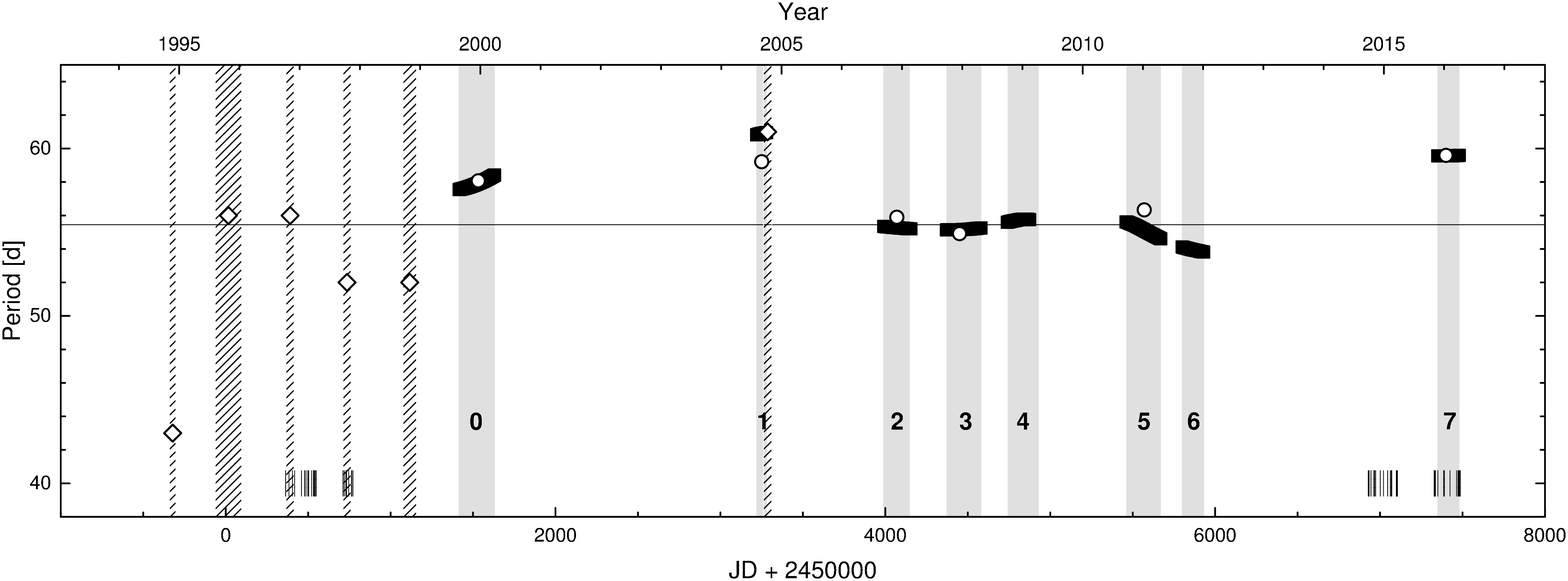}
 \caption{Long-term variability of V501\,Aur. The span of observations within each season is denoted by hatched pattern
          \citep[data from][]{grankin08} and light grey areas (this work).
          Period estimates derived per single season are plotted with diamonds \citep[data from][]{grankin08} and circles (this work). 
          The horizontal line represents the average period resulting from the entire data set (55.45~days). 
          Maximum values of the $Z$-transform of the wavelet analysis are shown with black squares. Short vertical lines along the bottom denote 
          times of spectroscopic observations. See Section~\ref{longterm} for details. \label{fig:wwz-all}}
\end{figure*}

\subsection{Wavelet analysis}
\label{longterm}

\hyphenation{ana-ly-sis}

\begin{figure}
 \centering   
 \includegraphics[width=82mm,clip=4]{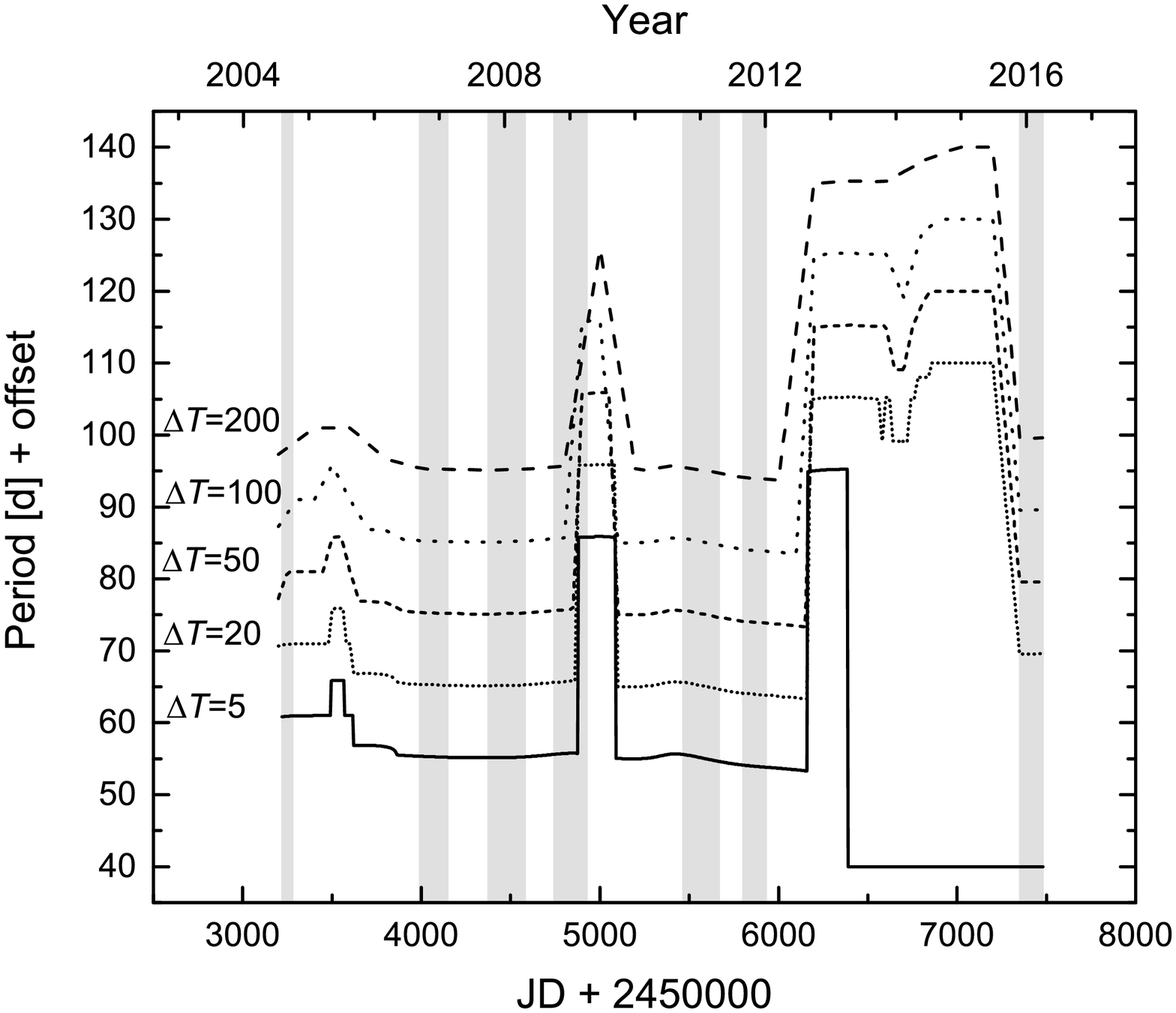}
 \caption{Comparison of different runs of our wavelet analysis. Individual runs are separated by an offset of 10 days. 
          Grey areas show the actual seasons with available data.}   
 \label{fig:wwz-comparison}
\end{figure}

To investigate the period variability further we have employed a standard time-frequency analysis with the Weighted Wavelet $Z$-Transform 
algorithm of \citet{Fos96}. This algorithm is also implemented in the {\tt VStar} package. We merged all seasons into a single data set and 
ran the search with a range of periods of 40--100 days, a period step of 0.1 day, and the so-called decay parameter (wavelet window) fixed at 
0.001 to get better resolution for period variations. We tested several time steps $\Delta T=200$, 100, 50, 20, and 5 days. 
Because the gaps in our data set are significantly larger than the span of the seasons, we considered it important to determine how the different 
binning affects the shape of the wavelet. We selected from the two-dimensional wavelet the maximum values of the amplitude in selected Julian date 
bins, and the results are displayed in Figure~\ref{fig:wwz-comparison}. The analysis with a time step of $\Delta T=5$ days diverged between 
Seasons~6 and 7, but all other time steps produced results that were similar. During date intervals with available data 
(grey areas in Figure~\ref{fig:wwz-comparison}) all runs produced the same results. We also tested a broader period interval, 
but the wavelet diverged to one of the border periods when encountering a large gap in the data set. When referring to the outcome of the wavelet 
analysis we use results from the run with date separation $\Delta T=20$ days. 

\begin{figure}
 \centering
 \includegraphics[width=82mm,clip=4]{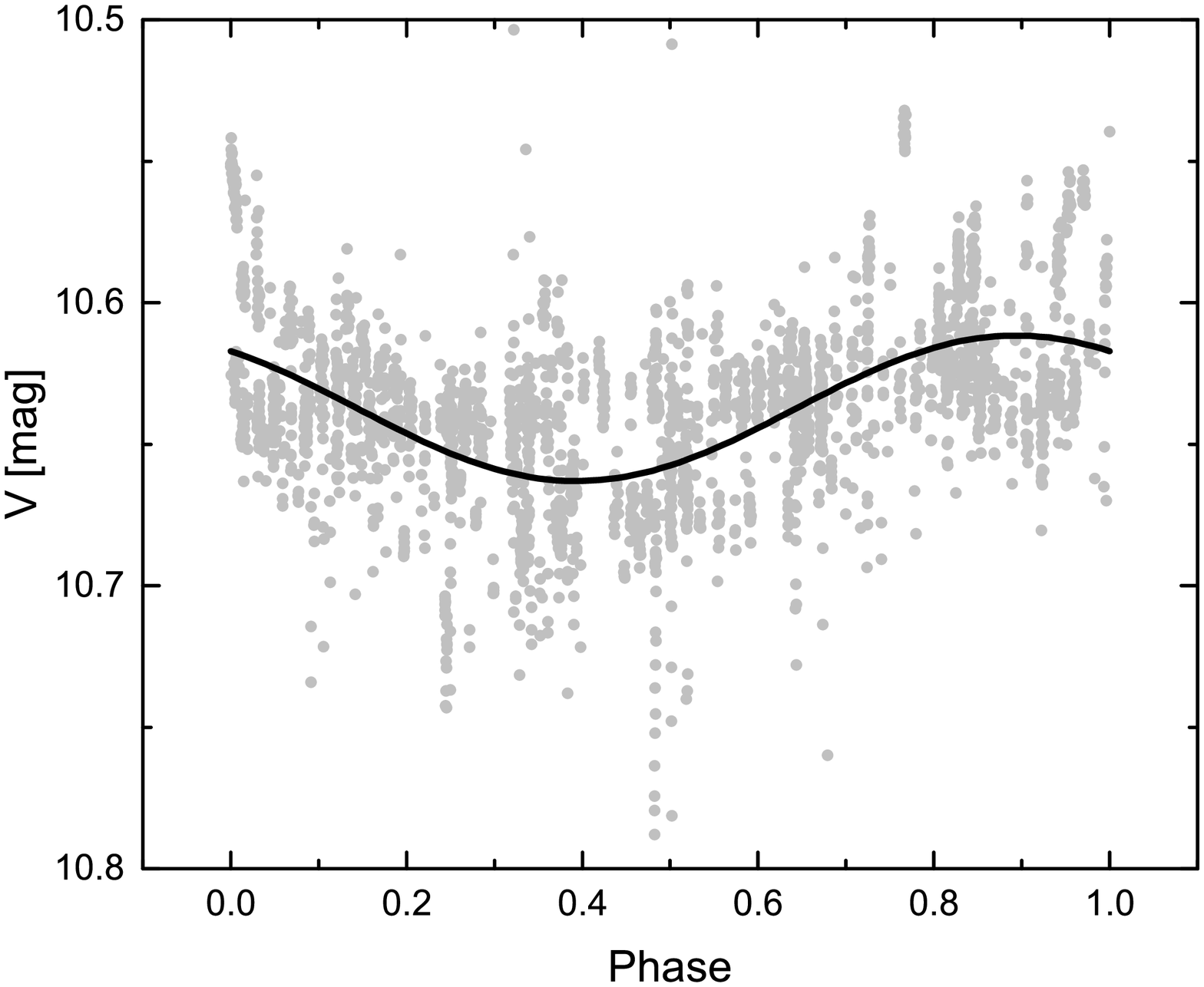}
 \caption{Phase diagram for all available data of V501\,Aur folded with the average period of $P=55.45$ days.}
 \label{fig:phaseall}
\end{figure}

We kept the maximum wavelet $Z$-values only for Julian date bins corresponding to actual data. 
The periods tabulated in Table~\ref{tab:seasons} are in a good agreement with the wavelet analysis. 
Since we were unable to find periods from the period analysis for Seasons~4 and 6, we provide a rough estimate 
as $\sum_{i=1}^{N} Z_{\rm max}/\sum_{i=1}^{N} i$ where $i$ runs through all $N$ bins in a given season. 
The wavelet in Season~4 changed its value abruptly from $\sim$55 days to $\sim$85 days toward the end of the data set. 
We have no explanation for this. This could be caused by an intrinsic change of variability similar to that in Season~3 
as discussed in the previous section. The results of the wavelet analysis and Fourier analysis (Section~\ref{period}) are 
summarized in Figure~\ref{fig:wwz-all}. Finally, a folded light curve using all the data and the average period of $P = 55.45$~days 
(Table~\ref{tab:seasons}) is shown in Figure~\ref{fig:phaseall}.

\section{Spectroscopic observations}
\label{spectroscopy}

Spectroscopic observations of V501\,Aur were carried out with three different instruments. We began at the CfA in October 1996 using the 
(now decommissioned) Digital Speedometer \citep[DS;][]{latham92} mounted on the 1.5m Tillinghast reflector at the Fred L. Whipple Observatory 
on Mount Hopkins (Arizona). Twenty-four spectra were recorded from September to November 1997. Some of these were used in the studies 
of \citet{wichmann00} and \citet{vanko15} cited in Section~\ref{intro}. The resolving power of this instrument was $R \sim 35\,000$, 
containing a single \'echelle order 45~\AA\ wide centered on the Mg~I triplet (5165.8--5211.2~\AA). Reductions were carried out with a dedicated 
pipeline, with the wavelength solution being set by exposures of a Thorium-Argon lamp before and after each science exposure. 
The velocity zero-point of this instrument was monitored with exposures of the dusk and dawn sky, and small run-to-run corrections 
were applied to the velocities described below as explained by \citet{latham92}. The SNRs of these observations range from 10 to 30 per 
resolution element of 8.5~$\kmps$. All spectra appear single-lined.

Nine additional spectra were gathered from October 2014 to February 2015 with the Tillinghast Reflector \'Echelle Spectrograph 
\citep[TRES;][]{furesz08} on the same telescope. This bench-mounted, fiber-fed instrument provides a resolving power of $R \sim 44\,000$ in 
51 orders over the wavelength range 3900--9100~\AA. SNRs at 5200~\AA\ ranged from 21 to 44 per resolution element of 6.8~$\kmps$. 
Reductions and wavelength calibrations followed a procedure similar to that described above. IAU radial velocity (RV) standards were 
observed each night to monitor the velocity zero point.

Between September 2014 and April 2016 we obtained a further 25 spectra with the 60-cm Cassegrain telescope at the Star\'a Lesn\'a Observatory 
(G1 pavilion) using the fiber-fed \'echelle spectrograph eShel \citep{pribulla15}. The spectra consisting of 24 orders cover the wavelength range 
from 4150 to 7600~\AA. The resolving power of the spectrograph is $R=10\,000$--12\,000. 
The reduction of the raw frames and extraction of the 1D spectra have been described by
\citet{pribulla15}. The wavelength reference system, as defined by the preceding and following Thorium-Argon exposures, 
was stable to within 0.1~$\kmps$. The SNRs of the spectra at 5500~\AA\ range from 11 to 42 (see Table~\ref{tab03}).

\subsection{Radial velocities}
\subsubsection{CfA data}
\label{sec:cfadata}

Radial velocities from the CfA/DS and CfA/TRES spectra were obtained
by cross-correlation using the IRAF task {\tt XCSAO}
\citep{kurtz92}. Only the order centered on the Mg~I triplet was used
for both sets of spectra, for consistency. An appropriate template was
selected as described later from an extensive library of pre-computed
synthetic spectra \citep[see][]{nordstrom94, latham02}, with the
following parameters that are close to the final values adopted in
this work: effective temperature $T_{\rm eff} = 4750$~K, surface
gravity $\log g = 2.5$, rotational velocity 25~$\kmps$, and solar
metallicity. Radial velocities from the two instruments were placed on
the same reference frame to well within 0.1~$\kmps$. Results in the
heliocentric frame are listed in Table~\ref{tab03}.

\begin{table*}
\caption{Heliocentric radial velocities of V501\,Aur from CfA and G1. 
         Also listed are the estimated uncertainties and the SNRs at 5200~\AA\ (DS and TRES) or 5500~\AA\ (eShel).
\label{tab03}}
\footnotesize  
\begin{center} 
\begin{tabular}{lrccl|lrccl}
\hline
HJD         & RV        & $\sigma$  & SNR & source & HJD	  & RV        & $\sigma$  & SNR & source\\ 
2\,400\,000+& [$\kmps$] & [$\kmps$] &     &        & 2\,400\,000+ & [$\kmps$] & [$\kmps$] &	&	\\
\hline 
50362.9153  & $-$13.8 &   1.0   & 11 & DS     &  56966.7880  & $-$23.3  &   0.8   & 24 & TRES \\
50383.8742  &    12.8 &   0.8   & 12 & DS     &  56972.8815  &  $-$8.5  &   0.4   & 33 & TRES \\
50404.7680  & $-$28.6 &   0.9   & 12 & DS     &  57001.7760  &  $-$2.6  &   0.4   & 41 & TRES \\
50417.7154  & $-$41.4 &   0.9   & 11 & DS     &  57020.8238  & $-$40.1  &   0.5   & 21 & TRES \\
50459.5785  &     2.3 &   1.1   & 10 & DS     &  57045.7911  &     0.9  &   0.4   & 38 & TRES \\
50477.7025  & $-$36.6 &   0.8   & 13 & DS     &  57062.8015  &    11.0  &   0.5   & 22 & TRES \\
50478.5889  & $-$36.7 &   0.9   & 15 & DS     &  57068.2735  &     1.9  &   0.6   & 22 & eShel \\
50486.6598  & $-$39.6 &   1.2   & 13 & DS     &  57071.3213  &  $-$4.5  &   0.4   & 34 & eShel \\
50497.6765  & $-$18.8 &   0.8   & 11 & DS     &  57098.3689  & $-$35.4  &   0.6   & 21 & eShel \\ 
50503.5285  &  $-$5.6 &   1.1   & 11 & DS     &  57099.2818  & $-$33.7  &   0.7   & 19 & eShel \\
50521.5218  &    12.7 &   1.5   & 12 & DS     &  57102.2896  & $-$27.3  &   0.5   & 28 & eShel \\
50532.5676  &  $-$4.5 &   1.2   & 11 & DS     &  57105.2564  & $-$20.6  &   0.6   & 22 & eShel \\
50536.6212  & $-$15.2 &   0.9   & 11 & DS     &  57327.5888  &    11.2  &   0.3   & 38 & eShel \\
50541.5795  & $-$25.9 &   1.1   & 10 & DS     &  57330.5687  &    12.9  &   0.7   & 19 & eShel \\
50543.5327  & $-$31.2 &   0.8   & 11 & DS     &  57331.6593  &    13.1  &   0.3   & 42 & eShel \\
50549.5373  & $-$39.8 &   1.1   & 11 & DS     &  57332.6283  &    12.7  &   0.5   & 26 & eShel \\
50710.8428  &  $-$3.4 &   0.6   & 18 & DS     &  57335.6321  &    11.9  &   1.2   & 11 & eShel \\
50718.8748  &    10.2 &   0.8   & 17 & DS     &  57350.4565  & $-$13.1  &   0.5   & 24 & eShel \\ 
50728.7420  &    13.3 &   0.8   & 14 & DS     &  57350.5393  & $-$13.6  &   0.4   & 31 & eShel \\
50732.8548  &     7.6 &   0.7   & 16 & DS     &  57385.3840  &  $-$8.9  &   0.9   & 15 & eShel \\
50745.7864  & $-$20.8 &   0.7   & 16 & DS     &  57390.4150  &     2.9  &   1.1   & 12 & eShel \\
50762.9408  & $-$37.1 &   0.5   & 30 & DS     &  57424.3079  & $-$25.9  &   0.7   & 20 & eShel \\
50764.7501  & $-$36.5 &   0.9   & 12 & DS     &  57464.2662  &     8.6  &   0.6   & 21 & eShel \\
50771.7938  & $-$23.0 &   0.9   & 12 & DS     &  57472.2826  &    11.5  &   0.7   & 20 & eShel \\
56928.6225  &     4.9 &   0.6   & 22 & eShel  &  57476.2661  &     9.3  &   0.5   & 29 & eShel \\
56930.6061  &     2.7 &   0.5   & 27 & eShel  &  57477.2859  &     8.8  &   0.6   & 23 & eShel \\ 
56933.0178  &  $-$3.3 &   0.4   & 34 & TRES   &  57479.2961  &     5.4  &   0.8   & 16 & eShel \\
56944.9627  & $-$29.9 &   0.5   & 34 & TRES   &  57484.2870  &  $-$5.8  &   0.6   & 20 & eShel \\
56961.9875  & $-$33.1 &   0.4   & 44 & TRES   &  57486.2934  &  $-$8.8  &   0.6   & 21 & eShel \\
\hline				   
\end{tabular}
\end{center} 
\end{table*}  

Preliminary orbital fits to these radial velocities indicated a large
mass function, suggesting a massive secondary. Consequently, we made
an effort to detect this star in these spectra employing the
two-dimensional cross-correlation technique TODCOR \citep{zucker94}
and a broad range of trial templates for the secondary star. We found
no convincing evidence of a second set of lines down to a flux ratio
of approximately 0.05. The Mg triplet region was used for two principal reasons:
(i) experience has shown that this order has the most information on the
radial velocity (large number of strong metallic lines); (ii) our synthetic 
template spectra are optimized for, and only cover that region. 
They are based on a line list that was painstakingly tuned to match real 
stars in this wavelength region.

An additional attempt to detect the secondary component was made using
the broadening function technique \citep{rucinski92}. Portions of the
TRES spectra in the range 4900--5500~\AA\ were first deconvolved using
several K-type slowly rotating dwarfs as templates.  The resulting
broadening functions (BFs) always showed a single component rotating
at $\vsini = 24.7\pm0.7~\kmps$ (the primary star).  Use of a K2V
template (HD\,3765) revealed a weak additional component with an
intensity ratio $L_2/L_1 \sim 0.01$. However, the radial velocity of
this component was found to be constant in the geocentric frame, and
likely results from telluric lines in the red part of the spectrum.
The detection limits from this exercise depend mainly on the unknown
projected rotational velocity of the secondary, and faint and rapidly
rotating stars are considerably more difficult to detect. If the
secondary rotates with the same projected velocity as the primary, the
detection limit is then estimated to be $L_2/L_1 \sim 0.05$--0.10.

The extracted BFs were inspected to see the presence of dark spots indicated
by the photometric wave. The BF changes are marginal and not conclusive.
For $R$ = 44\,000 we can resolve only 8 pixels across the rotational profile 
in the velocity space. To conclusively prove the presence of the spots,
more spectra  of a higher resolution and SNR would have to be taken 
within one orbital period.

\subsubsection{Star\'a Lesn\'a data}

Two approaches were used to derive the RVs from the G1 data: (i)
cross-correlation against the spectrum of V501\,Aur with the highest
SNR serving as a template, and (ii) the BF technique, with BFs
extracted using as a template the spectrum of HD\,65583 (K0V,
$\vsini=3.3\pm1.7~\kmps$, ${\rm [Fe/H]} = -0.70$). For the
cross-correlation analysis we avoided spectral regions affected by
strong telluric features and used only the wavelength ranges
4700--5860~\AA, 5970--6260~\AA, 6330--6860~\AA, and
6975--7130~\AA. For the BF extraction we used only the green-yellow
part of the spectrum from 4900 to 5500~\AA, including the Mg~I
triplet. Because the spectral resolution in this case is comparable to
the $\vsini$ of V501\,Aur, a simple Gaussian function was fitted to
both the cross-correlation functions (CCFs) and the BFs to determine
the velocities.

The RVs derived from BFs were shifted to the IAU system using $\RV=13.2~\kmps$ for HD\,65583 (Evans, 1979). 
The RVs obtained by cross-correlation were shifted to be consistent with the IAU system using the difference 
between the systemic velocities from preliminary orbital fits (the shift was $\Delta\RV=12.8~\kmps$).
The spectroscopic elements from these fits obtained for both cases (CCFs and BFs) were consistent 
within 1--2$\sigma$, but the RVs derived from the BFs resulted in a slightly smaller residual 
standard deviation from the orbital fit (0.47$~\kmps$ for BFs vs. 0.53$~\kmps$ for CCFs), and were thus 
adopted for the final orbital solution.

\subsubsection{Spectroscopic orbit}

\begin{figure}
\begin{center}
\includegraphics[width=82mm,clip=4]{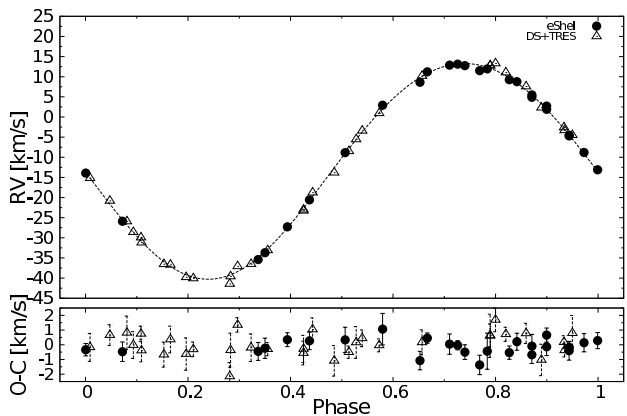}
\end{center}
\caption{Spectroscopic orbit of the primary component of V501\,Aur. 
         Phase 0.0 corresponds to periastron passage. Filled symbols represent radial velocities from G1, 
         and empty ones are from the CfA. For clarity the error bars are only shown for the residuals (bottom panel). 
\label{fig01}}
\end{figure}

Prior to performing a combined solution of the CfA and G1 observations
we tested the consistency of the RV reference systems by fitting
spectroscopic orbits to each data set separately. Only a small
difference in the systemic velocities was found ($\sim$0.3~$\kmps$),
so the data sets were combined without adjustments. The resulting
spectroscopic orbital elements for the merged data along with the
predicted times of spectroscopic conjunction are listed in
Table~\ref{tab03a}. A graphical representation of the measured radial
velocities and orbit model is shown in Figure~\ref{fig01}. The orbital
period is very well determined because of the 7123-day (nearly
20-year) time span of the velocities ($\sim$103 revolutions). The
orbital eccentricity is small but statistically significant, and
consistent between the CfA and G1 data sets.

\begin{table}
\caption{Spectroscopic orbital elements of the primary component of V501\,Aur based on radial-velocity measurements from CfA and G1. 
         The uncertainty is given in parentheses in units of the final digit. 
         $T_{\mathrm {min\,I}}$ and $T_{\mathrm {min\,II}}$ are the predicted times of primary 
         and secondary eclipse (spectroscopic conjunctions). 
         The columns present separate solutions from the CfA, G1, and combined data sets. 
         Heliocentric Julian dates (HJD) for periastron passage are given relative to HJD 2\,400\,000.
\label{tab03a}} 
\footnotesize  
\begin{center} 
\begin{tabular}{lccc}
\hline
Parameter                       &  CfA          &  G1            & Combined  \\
\hline
$P_{\mathrm {orb}}$ [d]         &  68.8347(14)  &  68.824(27)    & 68.8333(12)   \\
$e$                             &  0.020(7)     &  0.047(11)     & 0.030(5)      \\
$\omega$ [rad]                  &  3.2(3)       &  3.16(15)      & 3.13(15)      \\
$T_0$ [HJD]                     &  56\,955(3)   &  56\,954.1(17) & 56\,953.9(16) \\
$V_{\gamma}$ [$\kmps$]          &  $-$12.53(12) & $-$12.85(20)   & $-$12.71(8)   \\
$K_1$ [$\kmps$]                 &  27.04(19)    &  27.1(4)       & 26.78(13)     \\
$T_{\mathrm {min\,I}}$ [HJD]     &  56\,937(3)   &  56\,937.8(17) & 56\,937.5(16) \\
$T_{\mathrm {min\,II}}$ [HJD]    &  56\,971(5)   &  56\,970.1(25) & 56\,970.6(23) \\
$a_1 \sin i$ [au]               &  0.1710(12)   &  0.1715(26)    & 0.1693(8)     \\
$f(m)$ [$\Msun$]                &  0.142(3)     &  0.143(6)      & 0.1373(20)    \\
\hline
\end{tabular}   
\end{center}    
\end{table}

\subsection{Atmospheric parameters}

The main atmospheric properties of V501\,Aur, $T_{\rm eff}$, $\log g$,
and metallicity, were determined in several different ways. A first
method used the {\tt iSpec} software \citep{iSpec} applied to two of
the TRES spectra with the highest SNR, taken on October 31 and
December 10, 2014, and focusing on the spectral range 4800--5300
\AA. The effective temperature was determined from the line-strength
ratio of Cr\,I 4254~\AA\ to Fe\,I 4250~\AA\ and Fe\,I 4260~\AA\ (see
Digital Classification Spectral Atlas\footnote{\tt
  https://ned.ipac.caltech.edu/level5/Gray/Gray\_contents.html}), using template
spectra with well-established spectral types of K2V (HD\,3765), K3V
(HD\,128165), K4V (HIP\,073182), and K5V (61~Cyg\,A). In order to
minimize the number of free parameters the projected rotational
velocity was held fixed at the value $v \sin i = 24.7~\kmps$
determined from the BF fitting. The resulting parameters were $T_{\rm
  eff} = 5130\pm 410$~K, ${\rm [M/H]} = 0.03\pm 0.29$, and $\log
g=3.0\pm 0.7$, which are somewhat uncertain.  The EQW ratio of the
spectral lines mentioned above showed the best match for the K2V
template.

A second determination was carried out by cross-correlating each of
the DS spectra against the library of synthetic spectra, to find the
best match as a function of the template parameters $T_{\rm eff}$,
$\log g$, and $v \sin i$. Solar metallicity was assumed. The best
match was determined from the peak value of the CCF averaged over all
24 spectra. Because of degeneracies caused by the narrow
45~\AA\ window, it is generally difficult to establish the temperature
and $\log g$ at the same time from these spectra: lowering the
temperature and at the same time lowering the $\log g$ results in a
fit of similar quality, particularly when the SNR is low. We applied
the same approach to the nine TRES spectra, which have better SNR and also
a wider wavelength coverage around 100~\AA\ in the Mg\,I order. The
resulting average parameters for V501~Aur from these determinations
are $T_{\rm eff} = 4900 \pm 150$~K, $\log g = 2.7 \pm 0.6$, and $v
\sin i = 27\pm 2~\kmps$.  Because the temperature is also sensitive to
the adopted metallicity, which is not known very well, we have chosen
to assign a more conservative $T_{\rm eff}$ uncertainty of 250~K.

We also carried out an independent analysis of eight of our best TRES
spectra using the Stellar Parameters Classification tool
\citep[SPC;][]{buchhave12, buchhave14} obtaining the following
results: $T_{\rm eff} = 4685 \pm 100$~K, $\log g = 2.27 \pm 0.24$,
${\rm [M/H]} = -0.42 \pm 0.11$, and $v \sin i = 28.2 \pm 2.0~\kmps$.
These estimates are subject to similar degeneracies as mentioned
above.

The low surface gravity characteristic of giant stars was
independently tested using the luminosity-sensitive line ratio between
the Y\,II~4376~\AA\ and Fe\,I~4383~\AA\ lines, as recommended in the
Digital Classification Spectral Atlas for G8 stars. A spectrum close
to the lines in question was synthesized using {\tt iSpec} for several
values of the surface gravity. The comparison of the observed and
synthetic spectra supports the classification of the visible component
of V501\,Aur as a sub-giant or a giant. However, discriminating
between values of the surface gravity in the range $\log g = 1$--3 is
difficult.

The EQW of the Li\,I~$\lambda$6707 line was measured in each of our nine
high-resolution TRES spectra, which show the feature clearly. The result,
$90 \pm 6$~m\AA, is significantly lower than the value of 138~m\AA\ 
reported by \cite{wichmann00}, although the latter is based on
spectra of low SNRs (10--15) and according to those authors may have an
uncertainty of up to 40~m\AA. Unfortunately the line
cannot be measured reliably in our medium-dispersion spectra with
eShel because the blaze function has its minimum close to the location
of this feature. Finally, we examined the H$\alpha$ line in both our
TRES and eShel spectra and found that it is never seen in emission
over the year and a half of observation with those two instruments,
contrary to what was reported originally by \citet{wichmann96}. The
average EQW from our TRES spectra is $0.97 \pm 0.07$~\AA, similar to
the measure by \citet{martin99}.

\section{Interstellar extinction and distance}
\label{extinction}

The determination of the extinction towards V501\,Aur is complicated
by possible dust clumps in the young star-forming region against which
the star is projected. It has been shown that the EQW of the Na\,I~D1
line at 5896~\AA\ allows for a useful estimate of the $E(B-V)$ reddening.
For example, \citet{pozn12} provided the empirical relation
\begin{equation}
 \label{eq:ew2ebv}
  \log E(B-V)=2.47\times {\rm EQW}_{\mathrm {Na\,I\,D1}}-1.76\pm 0.17
\end{equation}
with EQW$_{\mathrm {Na\,I\,D1}}$ expressed in units of \AA. We used four of our
nine TRES spectra in which the stellar and interstellar components of the
D1 line are sufficiently separated, and fit double-Gaussians to the
normalized spectra, obtaining an average EQW$_{\mathrm
  {Na\,I\,D1}}=353$~m\AA. The equation above then leads to $E(B-V) =
0.13 \pm 0.06$. Alternatively, the look-up table provided by
\citet{muzwi97} results in a similar value of $E(B-V) = 0.15$. We note,
however, that the interstellar D1 line is saturated in our spectra,
implying that the above reddening estimates are only lower limits.

The {\it Gaia\/} mission \citep{GAIA16a} has recently provided a
measure of the trigonometric parallax of V501\,Aur as $\pi = 1.258 \pm
0.397$~mas, corresponding nominally to a distance range of about
600--1160~pc.  Despite the significant uncertainty, this indicates the
star is not a member of the Taurus-Auriga star-forming region, as has
been claimed before (see Section~\ref{intro}) but lies instead in the
background. The new distance estimate along with the 3D dust map from
Pan-STARRS \citep{green15} allow for a more reliable measure of the
reddening to be obtained. At the location of V501\,Aur, only 7 degrees
south of the Galactic plane, the result is $E(B-V) = 0.54 \pm 0.04$.

This reddening estimate is potentially useful to derive an independent
measure of the effective temperature from colour indices, for
comparison with the various spectroscopic determinations.
Unfortunately, however, colour measurements in the optical such as the
$B-V$ index differ considerably depending on the source, possibly due
to variability, or are rather uncertain, and are therefore unreliable.
For example, the {\it Tycho-2\/} measurements \citep{hog00} yield $B-V
= 1.29 \pm 0.21$ after transformation to the Johnson system, while the AAVSO Photometric
All Sky Survey \citep[APASS;][]{henden14} reports $B-V = 1.71 \pm
0.13$. Intermediate values have been given by others, such as $B-V =
1.62$ \citep{grankin08} and $B-V = 1.47 \pm 0.27$
\citep{kharchenko09}.

Near-infrared colours are less affected by reddening. For the 2MASS
$J-K_s$ index, we find using our above $E(B-V)$ estimate that $E(J-K_s)
= 0.523 \times E(B-V) = 0.282 \pm 0.021$ \citep[][Table~3]{cardel89}.
Applying this correction to the apparent $J-K_s$ colour of V501\,Aur,
$J-K_s = 0.948 \pm 0.031$, results in a de-reddened value of
$(J-K_s)_0 = 0.666 \pm 0.037$ in the 2MASS system.  Tabulations for
giant stars by \citet{ducati01} and \citet{gray92} then yield an
effective temperature range of 4750--4950~K, consistent with our
earlier spectroscopic estimates.

\section{The nature of the system}
\label{properties}

\subsection{Membership in the Taurus-Auriga SFR}
\label{membership}

As described in Section~\ref{intro}, V501\,Aur has been claimed to
belong to the central area of the Taurus-Auriga SFR \citep{wichmann96,
  frink97, wichmann00} mainly on the basis of its detection in X-rays
(ROSAT), the presence of the Li\,I~$\lambda$6707 line in absorption, the
late spectral type (K2), the reported emission in H$\alpha$, and a
proper motion apparently consistent with the average value for the
complex. Further properties of the star have been inferred in these
studies from the assumption of a distance of 140~pc to the
SFR. However, much of the evidence for a pre-main-sequence status is
somewhat circumstantial.

Information gathered since, as well as observations reported here,
paint a rather different picture of the nature of the system that we
now describe. The principal facts supporting the new interpretation
are the following, in order of relevance:

\begin{itemize}

\item The first public data release (DR1) from the {\it Gaia\/}
  mission has supplied the trigonometric parallax of V501\,Aur as $\pi
  = 1.258 \pm 0.397$~mas, corresponding a nominal distance of $d =
  795^{+366}_{-191}$~pc, or a range of approximately 600--1160~pc.
  This appears to rule out a membership in the Taurus-Auriga SFR
  \citep[$ d = 115$--156 pc, according to][]{grankin13}, and places
  the star in the background. The formal parallax uncertainty given
  above excludes a component of systematic error ($\sim$0.3~mas) that
  the {\it Gaia\/} Collaboration has recommended be factored into the
  total uncertainty \citep{GAIA16a}. If added quadratically to the
  internal error, it yields a slightly larger distance range of
  570--1300~pc, still ruling out a membership in Taurus.

\item An independent estimate of the distance may be derived if we
  interpret the photometric wave as being caused by rotational
  modulation due to cool photospheric spots on the visible component
  (see next section).  In that case, the measured projected rotational
  velocity along with the measured photometric period yield a lower
  limit to the radius of the star, $R \sin i = \frac{P_{\rm rot}}{2\pi} v \sin i$, 
  which combined with an effective
  temperature estimate provides a lower limit to the bolometric
  luminosity. This, in turn, can be used to infer a lower limit to the
  distance. An absolute lower limit to $R$ is obtained by using our
  smallest estimate of $v \sin i = 24.7~\kmps$ and the shortest
  photometric period from Table~\ref{tab03a}, $P_{\rm rot} \approx
  53.9$~days, resulting in $R \sin i \approx 26.3~\Rsun$. This already
  seems considerably larger than expected for a PMS star of any
  reasonable mass. Our coolest temperature estimate, $T_{\rm eff} =
  4685$~K, then implies that the luminosity of the star must be at
  least 300~$\Lsun$. If we assume no extinction, and adopt 
  the most recent value (and one of the brightest) of the apparent visual magnitude available for
  V501\,Aur \citep[$V = 10.57 \pm 0.12$;][]{henden14} along with a
  bolometric correction of $BC_V = -0.48 \pm 0.10$ for a star of this
  temperature \citep{flower96, torres10}, we obtain an absolute lower
  limit to the distance of $2050 \pm 180$~pc.
  Using $V = 10.88 \pm 0.08$ \citep[the faintest measurement in
  literature;][]{hog00} we obtain distance of $2350 \pm 200$~pc. Accounting for
  extinction according to $A_V = 3.1\,E(B-V)$ reduces this to $950 \pm
  100$~pc and $1100 \pm 100$~pc, respectively; which is consistent with the {\it Gaia\/} estimate
  and still much larger than the distance to Taurus-Auriga. This supports
  the conclusion that V501\,Aur is far behind the SFR.

\item Our orbital solution for V501\,Aur (Table~\ref{tab03a}) yields a
  systemic velocity of $-12.70 \pm 0.08~\kmps$ that is inconsistent
  with the mean radial velocity of the Taurus-Auriga SFR \citep[$+9.8
    < {\rm RV} < +17.5~\kmps$;][]{mooley13}, arguing strongly against
  a membership.

\item The EQW of the Li\,I~$\lambda$6707 line as measured from our nine
  TRES spectra, $90 \pm 6$~m\AA, is lower than is typical for a very
  young object of this temperature \citep[see Fig.~2
    of][]{wichmann00}, making the PMS status unlikely. We also do not
  find H$\alpha$ to be in emission in our spectra, as is usually the
  case for WTTS. Compared to other evolved stars the lithium abundance
  of V501~Aur is high, but not exceptional \citep[see][]{rand99}. 
  \citet{drake02} list a few rapidly-rotating giants showing
  large Li abundances and enhanced activity (indicated by X-ray luminosity 
  and Ca II emission) similar to V501~Aur.

\item Most spectroscopic estimates of the surface gravity of the
  visible star from our analysis ($\log g = 2.3$--3.0) are
  considerably lower than expected for very young objects, again
  pointing towards the view that V501\,Aur is an evolved star rather
  than a PMS star.

\item The measured proper motion of V501\,Aur is rather small and thus
  not a very reliable indicator of a membership in Taurus-Auriga.
  \citet{frink97} reported $(\mu_\alpha \cos\delta, \mu_\delta) =
  (-2.0 \pm 4.4, -20.0 \pm 4.4)$ mas~yr$^{-1}$, and considered this to
  be consistent with the mean proper motion for the SFR of $(+4.0,
  -18.7)$ mas~yr$^{-1}$. More recent estimates from {\it Tycho-2\/}
  give a smaller total motion of $(-5.4 \pm 2.3, -11.9 \pm 2.3)$
  mas~yr$^{-1}$, and {\it Gaia}/DR1 lists the star as having an even
  smaller motion of $(-8.3 \pm 2.6, -3.4 \pm 1.9)$
  mas~yr$^{-1}$. Given that the proper motions of the true members
  show considerable scatter, and that there is also a significant
  difference between the central parts of Taurus-Auriga, ($+2.4,
  -21.2$) mas~yr$^{-1}$, and the southern region, ($+10.1, -9.8$)
  mas~yr$^{-1}$, the proper motion criterion for V501\,Aur is largely
  inconclusive.

\end{itemize}

The above evidence strongly supports the notion that V501\,Aur is a
background giant, rather than a young member of the Taurus-Auriga SFR.

\subsection{Photometric variability}

Given the binary nature of V501\,Aur, we searched for eclipses near
the predicted times of spectroscopic conjunction but found none. Here
we discuss other mechanisms that might explain the quasi-periodic
brightness changes (see Sections~\ref{period} and \ref{longterm}).

Ellipsoidal variability caused by distortions in the large primary
star seem rather unlikely in view of the long orbital period. 
The modulation would also be strictly periodic with a
period exactly half that of the orbital period, $P/2 \approx 34.4$~days. 
A detailed search in the vicinity of this value using the more numerous superWASP 
data with the best temporal coverage, as well the entire photometric data set, showed
no convincing evidence of a signal at this period. Numerical simulations of
the expected amplitude of such an effect based on approximate stellar
properties for the primary star and the companion as inferred in the next
section indicate a peak-to-peak variation of only about 1\% in the $V$
band. This is at the level of the noise in the observations, which may
explain the non-detection. It is also much smaller than the brightness
changes actually observed in V501\,Aur, so it cannot be the principal cause
of those variations. 

Pulsation, either in the primary or less likely in
the unseen secondary, is another possibility. Our period analysis
shows no coherent signals in the power spectrum. The long timescale of
the variations ($\sim$55~days) and the small colour changes
($\Delta(V-I) \approx 0.05$~mag) are consistent with the
characteristics of some semi-regular variables \citep[specifically
with those of the SRd type; see the GCVS,][]{gcvs}, although the
amplitude of the variations in V501\,Aur (0.013--0.082~mag) is smaller
than is typical for such objects, and the rapid rotation is also
unusual for this class.

The overall features of the photometric variability seem most
consistent with modulation from photospheric spots on the visible
component, perhaps appearing at different latitudes in a star with
differential rotation. The observations show that the amplitude of the
variations decreases toward the red, which is consistent with the spot
scenario. That the star is active seems supported by the strong X-ray
emission (ROSAT). Further evidence for spots could be obtained in
principle through further analysis of the BFs, or trailing spectra with
higher SNR and high spectral resolution. More quantitative modeling of
the spots may require multi-colour photometry and considerably better
time sampling than we presently have.

\subsection{Physical parameters}

The lack of eclipses in V501\,Aur prevents a direct determination of
the fundamental properties of the components, such as the masses and
radii, from our spectroscopic and photometric observations. As a
result, the nature of the secondary component is unclear. The large
velocity semi-amplitude of the primary ($K_1 = 26.8~\kmps$) results in
a fairly large mass function $f(m) = 0.1373 \pm 0.0002~\Msun$ and a
correspondingly large minimum mass for the secondary, $M_2\sin i =
0.516 (M_1 + M_2)^{2/3}~\Msun$. It is somewhat surprising, therefore,
that we see no trace of the secondary lines in our spectra despite our
attempts at detection (Section~\ref{sec:cfadata}). One possibility is
that the secondary is itself a close binary system. On the other hand,
if the primary is a giant star and the secondary a normal
main-sequence star, this would naturally explain our lack of detection
as the brightness contrast would be very unfavourable, particularly if
the secondary were to be rotating rapidly. This explanation would seem
to be supported by the arguments in Section~\ref{membership}.

We explored this scenario further by using stellar evolution models
from the MESA series \citep{mesa11, mesa13, mesa15}, seeking to match
all observational constraints from our orbital fit and our
spectroscopic analysis with a single isochrone placing the primary
star either on the giant branch or near the clump, and the secondary
star on the main sequence. We restricted the parameter space to $2.1 <
\log g < 3.3$ and $4585~{\rm K} < T_{\rm eff} < 5150$~K (following
from our spectroscopic CCF and SPC estimates and their 1$\sigma$
uncertainties), as well as $R > 26.3~\Rsun$
(Section~\ref{membership}). Solar metallicity was assumed.

\begin{figure}
\begin{center}
\includegraphics[width=\columnwidth,clip=4]{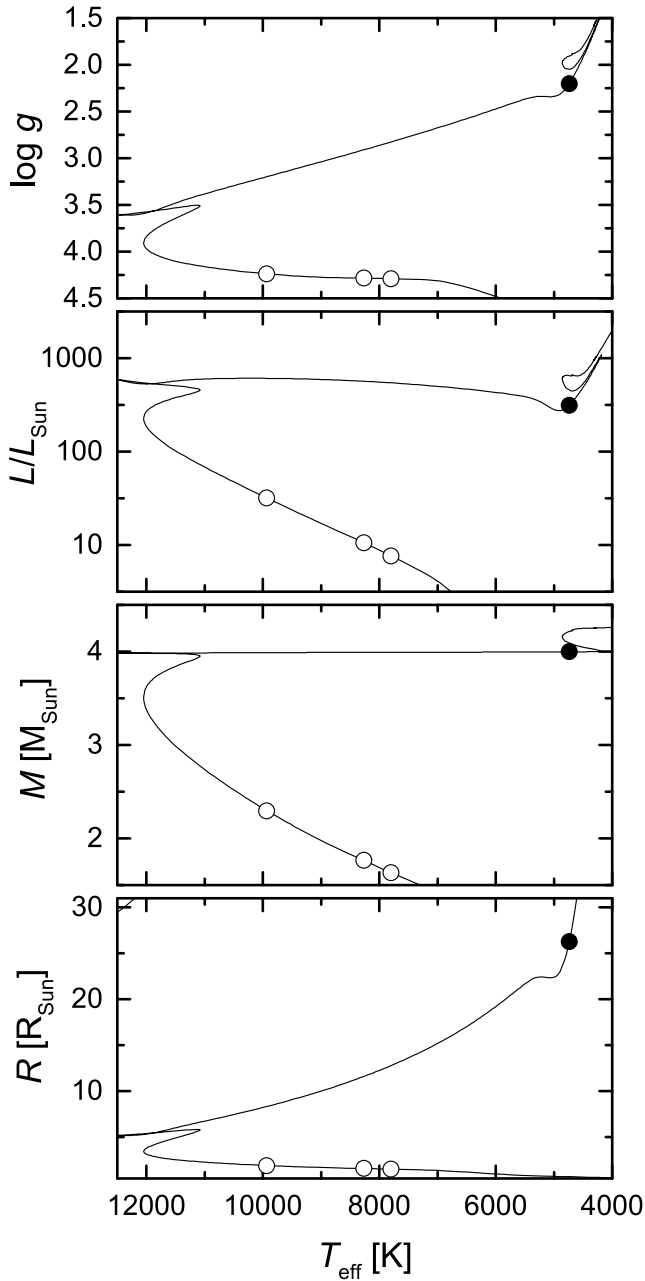}
\end{center}
\caption{MESA model isochrone for an age of 178~Myr and solar metallicity. 
         The filled circle near the top represents the primary component ($M_1 = 4~\Msun$), 
         and the open circles correspond to three possible secondary masses (2.30, 1.77, and 1.63~$\Msun$) 
         for representative inclination angles of 50, 70 and 90 degrees, respectively. 
\label{fig:lubos}}
\end{figure}

A satisfactory solution was found for a primary star at the bottom of
the giant branch with approximate parameters $M = 4~\Msun$, $R =
26.6~\Rsun$, $\log g = 2.19$, $T_{\rm eff} = 4727$~K, $L = 313~\Lsun$,
and an age of $\log({\rm age}) = 8.25$. These properties and the
best-fitting isochrone are shown in Figure~\ref{fig:lubos}.

The mass of the secondary star for a given orbital inclination angle
may be found from the primary mass and the spectroscopic mass
function. We mark in Figure~\ref{fig:lubos} three such values
corresponding to representative inclination angles of 50, 70, and 90
degrees, which give secondary masses of 2.30, 1.77, and 1.63~$\Msun$,
respectively\footnote{Note that an inclination angle of $90^{\circ}$
  is actually ruled out by the lack of eclipses (see previous
  section).}. Other properties of the secondary inferred from the
best-fitting isochrone are listed in Table~\ref{tabsec}. These results
would suggest that the secondary star is no later than spectral type
late A, and may be earlier if lower inclination angles are assumed.
The predicted luminosity ratios $L_2/L_1$ in Table~\ref{tabsec} are
seen to be fairly low, and this, compounded with the likely rapid
rotation expected for an A-type star, would explain the non-detection
of the secondary in our spectra.

\begin{table}
\caption{Estimated absolute parameters of the secondary component for three representative inclination angles $i$ = 50, 70 and 90 degrees. 
The table lists both bolometric and visual light ratios.
\label{tabsec}} 
\footnotesize  
\begin{center} 
\begin{tabular}{lccc}
\hline
$i$ [deg]    &   50 &    70 &   90 \\
\hline
$M~[\Msun]$  & 2.30 &  1.77 & 1.63 \\
$L~[\Lsun]$  &   32 &    13 &  7.6 \\
$R~[\Rsun]$  &  1.9 &  1.60 &  1.5 \\
$T_{\rm eff}$ [K]  & 9950 &  8260 & 7800 \\
$L_2/L_1$ (bol)   & 0.10 &  0.041 &  0.024 \\
$L_2/L_1$ ($V$)   & 0.11 &  0.055 &  0.033 \\
\hline
\end{tabular}
\end{center}
\end{table}

An important information on the nature of V501~Aur is provided by the observed X-ray flux and resulting X-ray luminosity.
Using the RASS count rate, 2.8(9)$\times 10^{-2}$ ct~s$^{-1}$, the X-ray luminosity for the 1-$\sigma$ range of the \textit{Gaia} distance is from 
8.7$\times 10^{31}$ erg~s$^{-1}$ to 3.2$\times 10^{32}$ erg~s$^{-1}$. This is three orders more than
a typical X-ray luminosity of late-type red giants, 10$^{28}$ - 10$^{29}$ erg~s$^{-1}$ \citep{maggi90}. The identification of the X-ray source with V501~Aur is, however, questionable.
The position offset of the X-ray source is 19 arcsec and the RASS beam size about 30 arcsec. Hence, observations with better resolution are needed prior to any interpretation.

\section{Discussion and final remarks}
\label{conclusion}

Our new photometric and spectroscopic observations of V501\,Aur have
enabled us to revisit its properties in the context of previous claims
that it is a WTTS belonging to the Taurus-Auriga star-forming region.
Our spectroscopic observations reveal it to be a slightly eccentric,
single-lined spectroscopic binary with an orbital period of about
68.8~days and a fairly massive unseen companion. No signs of eclipses
are observed.

We have carried out a detailed investigation of the long-term
photometric variability of V501\,Aur using our own new data augmented
with observations from several other sources, giving a total time span
of two decades. The changes in the previously known photometric wave
(in both period and amplitude), which has an average period of
$\sim$55~days distinctly shorter than the orbital period, and the
concomitant colour variations we have measured, suggest that the
variability is caused by photospheric spots appearing at different
stellar latitudes on a differentially rotating star.  The long period
of the photometric wave and the fairly rapid rotation of the visible
component ($\sim 25~\kmps$) imply a large radius for the star of $R >
26.3~\Rsun$.

Such a large size is supported by our spectroscopic estimates of the
surface gravity giving low values of $\log g =2.3$--3.0. This strongly
suggests it is an evolved star that must be far behind the
Taurus-Auriga SFR. The systemic radial velocity of V501~Aur, $-12.70
\pm 0.08~\kmps$, is also inconsistent with the mean radial velocity of
known members of Taurus-Auriga.  Furthermore, our new measurements of
the Li\,I~$\lambda$6707 line show it to be much weaker than is typical for
a young T-Tauri star, and the H$\alpha$ line is not seen in emission,
as would be expected. The distance of about 800~pc now known from the
trigonometric parallax appearing in the first {\it Gaia\/} data
release conclusively rules out a membership in the SFR. An independent
distance estimate based on the minimum size of the star ($R \sin i$),
its temperature, the apparent brightness, and a measure of
interstellar extinction are perfectly consistent with the direct
measurement. Thus, the previous classification of V501\,Aur as a WTTS
belonging to Taurus-Auriga is overwhelmingly {\it not\/} supported by
available observations.

Instead, the scenario that emerges for V501\,Aur, aided by a
comparison with stellar evolution models that succeed in matching all
observational constraints, is one in which it is a background,
non-eclipsing spectroscopic binary projected onto the Taurus-Auriga
SFR, with a luminous, spotted, and fairly rapidly rotating giant star
as the primary, and a likely much more rapidly rotating early-type
star as the secondary. The estimated age of the system is roughly
180~Myr, according to the models. The unfavourable luminosity ratio of
such a configuration along with the rotationally broadened lines
expected for the secondary are sufficient to explain the non-detection
of that star in our spectra. Another possible scenario is that 
the secondary component is a close binary.

V501\,Aur may be related to a group of rapidly rotating giant stars
recently identified in the photometry from the {\it Kepler\/} mission
\citep{costa15}. Among them, we note that KIC~10293335 has properties
somewhat similar to those of V501\,Aur: $P_{\rm rot} = 55.96$~days
(from a Fourier analysis), $T_{\rm eff} = 4363$~K, $\log g = 2.45$, $R
= 21.29~\Rsun$, and $v \sin i = 12.7~\kmps$. Interestingly, there are
hints that this star may also be a (single-lined) spectroscopic
binary. Further study of V501\,Aur in the framework of a possible
connection to this interesting class of stars would benefit from a
detailed chemical analysis, including sensitive diagnostics of
evolution such as the CNO abundances and the $^{12}$C/$^{13}$C isotope
ratio to help pinpoint its evolutionary state.

\section*{Acknowledgments}

We are grateful to A. Berndt, M. Moualla, T. Eisenbei\ss, M. Hohle and T.
R\"oll for help in obtaining the observations of V501\,Aur at the University
Observatory Jena. This work has been supported by the projects VEGA~2/0143/14, APVV-15-0458
and partially supported by SAIA scholarship. GT acknowledges partial support for this work 
from NSF grant AST-1509375. MV and TP would like to thank the European Union
in the Framework Programme FP6 Marie Curie Transfer of Knowledge project MTKD-CT-2006-042514 for support.
We would like to acknowledge financial support from the Thuringian government (B 515-070 10)
for the STK CCD camera used in this project at the University Observatory Jena.
The work of SS has been supported in part by the RFBR grant No. 15-02-06178 and NSh-9670.2016.2.
SR acknowledges support from the People Programme (Marie Curie Actions) of the European Union's 
Seventh Framework Programme (FP7/2007-2013) under REA grant agreement No.
[609305]. This paper has made use of data from DR1 of the WASP data as provided by the WASP consortium, and the computing and 
storage facilities at the CERIT Scientific Cloud, reg.\ no.\ CZ.1.05/3.2.00/08.0144 which is operated by Masaryk University, 
Czech Republic. This work has also made use of data from the European Space Agency (ESA) mission {\it Gaia\/} 
(http://www.cosmos.esa.int/gaia), processed by the {\it Gaia\/} Data Processing and Analysis Consortium 
(DPAC, http://www.cosmos.esa.int/web/gaia/dpac/consortium). 
Funding for the DPAC has been provided by national institutions, in particular
the institutions participating in the {\it Gaia\/} Multilateral Agreement.
This article was created by the realisation of the project ITMS No.26220120029, based on the 
supporting operational Research and development program financed from the European Regional Development
Fund.

\label{lastpage}

\begin{thebibliography}{99}
\bibitem[\protect\citeauthoryear{Blanco-Cuaresma et al.}{2013}]{iSpec}
Blanco-Cuaresma, S., Soubiran, C., Jofr\'e, P., Heiter, U.
2013, arXiv 1312.4545
\bibitem[\protect\citeauthoryear{Bouvier et al.}{1997}]{bouvier97}
Bouvier J., Wichmann R., Grankin K., et al.
1997, \aap, 318, 495
\bibitem[\protect\citeauthoryear{Buchhave et al.}{2012}]{buchhave12}
  Buchhave, L.~A., Latham, D.~W., Johansen, A., et al.\ 2012, Nature,
  486, 375
\bibitem[\protect\citeauthoryear{Buchhave et al.}{2014}]{buchhave14}
  Buchhave, L.~A., Bizzarro, M., Latham, D.~W., et al.\ 2014, Nature,
  509, 593
\bibitem[\protect\citeauthoryear{Butters et al.}{2010}]{butters10}
Butters O.W., West R.G., Anderson D.R., et al.
2010, \aap, 520, 10
\bibitem[\protect\citeauthoryear{Cardelli et al.}{1989}]{cardel89}
Cardelli, J.A., Clayton, G.C., Mathis, J.S.
1989, \apj, 345, 245
\bibitem[\protect\citeauthoryear{Carkner et al.}{1997}]{carkner97}
Carkner L., Mamajek E., Feigelson, E., et al.
1997, \apj, 490, 735
\bibitem[\protect\citeauthoryear{Costa et al.}{2015}]{costa15}
Costa, A.D., Canto Martins, B.L., Bravo, J. P. et al.
2015, \apj, 807, L21
\bibitem[\protect\citeauthoryear{Cox}{2001}]{cox01}
Cox A.N. 2001, Allen's Astrophysical Quantities 4$^{\rm th}$ ed., Springer.
\bibitem[\protect\citeauthoryear{Daemgen et al.}{2015}]{daemgen15}
Daemgen S., Bonavina M., Jayawardhana R. et al.
2015, \apj, 799, 155
\bibitem[\protect\citeauthoryear{D'Antona \& Mazzitelli}{1994}]{dantona94}
D'Antona F., Mazzitelli I.
1994, ApJS, 90, 467  
\bibitem[\protect\citeauthoryear{Drake et al.}{2002}]{drake02}
Drake, N.A.,2 Ramiro de la Reza,1 Licio da Silva,1 and David L. Lambert3
\bibitem[\protect\citeauthoryear{Ducati et al.}{2001}]{ducati01}
Ducati, J.~R., Bevilacqua, C.~M., Rembold, S.~B., \& Ribeiro,
D.\ 2001, ApJ, 558, 309
\bibitem[\protect\citeauthoryear{ESA}{1997}]{esa97} 
European Space Agency
1997, The Hipparcos and Tycho Catalogues (ESA SP-1200), Noordwijk
\bibitem[\protect\citeauthoryear{Ferraz-Mello}{1981}]{FM81}
Ferraz-Mello S. 1981, \aj, 86, 619
\bibitem[\protect\citeauthoryear{Flower}{1996}]{flower96} Flower,
  P.~J.\ 1996, \apj, 469, 355
\bibitem[\protect\citeauthoryear{Foster}{1996}]{Fos96}
Foster G. 1996, \aj, 112, 1709
\bibitem[\protect\citeauthoryear{Frink et al.}{1997}]{frink97}
Frink S., R\"oser S., Neuh\"auser R., Sterzik M.F.
1997, \aap, 325, 613
\bibitem[\protect\citeauthoryear{F\H{u}r\'esz}{2008}]{furesz08}
F\H{u}r\'esz G.
2008, PhD thesis, Univ. Szeged
\bibitem[\protect\citeauthoryear{Gaia Collaboration}{2016}]{GAIA16a}
Gaia Collaboration, Brown A.G.A., Vallenari A., Prusti T., et al.
2016, \aap, special Gaia volume
\bibitem[\protect\citeauthoryear{Grankin et al.}{2008}]{grankin08}
Grankin K.N., Bouvier J., Herbst W., Melnikov S.Yu.
2008, \aap, 479, 827
\bibitem[\protect\citeauthoryear{Grankin}{2013}]{grankin13}
Grankin K.N.
2013, AstL, 39, 251
\bibitem[\protect\citeauthoryear{Gray}{1992}]{gray92} Gray,
D.~F.\ 1992, The Observation and Analysis of Stellar Photospheres,
Camb.~Astrophys.~Ser., Vol.~20, p.\ 430
\bibitem[\protect\citeauthoryear{Green et al.}{2015}]{green15}
Green, G.M., Schlafly, E.F., Finkbeiner, D.P. et al.
2015, ApJ, 810, 25
\bibitem[\protect\citeauthoryear{Henden et al.}{2014}]{henden14}
Henden, A., \& Munari, U.\ 2014, Contributions of the Astronomical Observatory Skalnate Pleso, 43, 518 
\bibitem[\protect\citeauthoryear{H\o g et al.}{2000}]{hog00}
H\o g E., Fabricius C., Makarov V.V., et al.
2000, \aap, 355, 27
\bibitem[\protect\citeauthoryear{Holdsworth et al.}{2014}]{holdsworth14}
Holdsworth D.L., Smalley B., Gillon M., et al.
2014, \mnras, 439, 2078
\bibitem[\protect\citeauthoryear{Kholopov et al.}{1985}]{gcvs}
Kholopov P.N., Samus N.N., Frolov M.S. et al.,
1985, General Catalogue of Variable Stars, 4th Edition, Nauka, Moscow
\bibitem[\protect\citeauthoryear{Kharchenko et al.}{2009}]{kharchenko09}
Kharchenko, N.~V., Piskunov, A.~E., R{\"o}ser, S., et al.\ 2009, \aap, 504, 681
\bibitem[\protect\citeauthoryear{Kurtz}{1992}]{kurtz92}
Kurtz, M.J., Mink, D.J., Wyatt, W.F. et al. 1992, in Astronomical Data Analysis Software and Systems I, ASP Conf. Ser., Vol. 25, eds. D.M. Worral, C. Biemesderfer, and J. Barnes, 432
\bibitem[\protect\citeauthoryear{Latham}{1992}]{latham92}
Latham, D.W. 
1992, in IAU Coll. 135, Complementary Approaches to Double
and Multiple Star Research, ASP Conf. Ser. 32, ed. H. A. McAlister \&
W. I. Hartkopf (San Francisco, CA: ASP), 110
\bibitem[\protect\citeauthoryear{Latham}{2002}]{latham02}
Latham, D.W., Stefanik, R.P., Torres, G. et al. 2002, AJ, 124, 1144
\bibitem[\protect\citeauthoryear{Maggio et al.}{1990}]{maggi90}
Maggio, A., Vaiana, G.S., Haisch, B.M., Stern, R.A., Bookbinder, J., 
Harnden, F.R., Rosner, R.
1990, \apj, 348, 253
\bibitem[\protect\citeauthoryear{Mart\'in \& Magazz\'u}{1999}]{martin99}
Mart\'in E.L., Magazz\'u A.
1999, \aap, 342, 173
\bibitem[\protect\citeauthoryear{Mooley}{2013}]{mooley13}
Mooley, K., Hillenbrand, L., Rebull, L., Padgett, D., Knapp, G.
2013, \apj, 771, 110
\bibitem[\protect\citeauthoryear{Mugrauer}{2009}]{mugi09}
Mugrauer M. 
2009, \an, 330, 419
\bibitem[\protect\citeauthoryear{Mugrauer \& Berthold}{2010}]{mugi10}
Mugrauer M., Berthold T.
2010, \an, 331, 449
\bibitem[\protect\citeauthoryear{Munari \& Zwitter}{1997}]{muzwi97}
Munari, U., Zwitter, U.
1997, \aap, 318, 269
\bibitem[\protect\citeauthoryear{Nguyen et al.}{2009}]{nguyen09}
Nguyen D.C., Jayawardhana R., van Kerkwijk M.H., et al.
2009, \apj, 695, 1648
\bibitem[\protect\citeauthoryear{Nordstr\"om et al.}{1994}]{nordstrom94}
Nordstr\"om B., Latham D.W., Morse, J. et al. 1994, A\&A, 287, 338
\bibitem[\protect\citeauthoryear{Parimucha \& Va\v{n}ko}{2015}]{parimucha15}
Parimucha \v{S}, Va\v{n}ko M.
2015, Living Together: Planets, Host Stars and Binaries, eds. by
Slavek M. Rucinski, Guillermo Torres, and Miloslav Zejda. ASP Conference
Series, 496, 309
\bibitem[\protect\citeauthoryear{Paxton et al.}{2011}]{mesa11}
Paxton, B., Bildsten, L., Dotter, A., et al. 
2011, ApJS, 192, 3
\bibitem[\protect\citeauthoryear{Paxton et al.}{2013}]{mesa13}
Paxton, B., Cantiello, M., Arras, P., et al. 
2013, ApJS, 208, 4
\bibitem[\protect\citeauthoryear{Paxton et al.}{2015}]{mesa15}
Paxton, B., Marchant, P., Schwab, J., et al. 
2015, ApJS, 220, 15
\bibitem[\protect\citeauthoryear{Pollacco et al.}{2006}]{pollacco06}
Pollacco D.L., Skillen I., Collier Cameron A., et al.
2006, PASP, 118, 1407
\bibitem[\protect\citeauthoryear{Poznanski et al.}{2012}]{pozn12}
Poznanski D., Prochaska J.X., Bloom, J.S.
2012, \mnras, 426, 1465
\bibitem[\protect\citeauthoryear{Pribulla et al.}{2015}]{pribulla15}
Pribulla T., Garai Z., Hamb\'alek \v{L}., et al. 
2015, \an, 336, 682
\bibitem[\protect\citeauthoryear{Randich et al.}{1999}]{rand99}
Randich, S., Gratton, R., Pallavicini, R., Pasquini, L., Carretta, E.
1999, \aap, 348, 487
\bibitem[\protect\citeauthoryear{Rucinski}{1992}]{rucinski92}
Rucinski, S.M. 1992, AJ, 104, 1968
\bibitem[\protect\citeauthoryear{Smalley et al.}{2011}]{smalley11}
Smalley B., Kurtz D.W., Smith A.M.S., et al.
2011, \aap, 535, A3
\bibitem[\protect\citeauthoryear{Torres}{2010}]{torres10}
Torres, G.
2010, \aj, 140, 1158
\bibitem[\protect\citeauthoryear{Va\v{n}ko et al.}{2015}]{vanko15}
Va\v{n}ko M., Torres G., Pribulla T., et al. 
2015, Living Together: Planets, Host Stars and Binaries, eds. by
Slavek M. Rucinski, Guillermo Torres, and Miloslav Zejda. ASP Conference
Series, 496, 262
\bibitem[\protect\citeauthoryear{Wichmann et al.}{1996}]{wichmann96}
Wichmann R., Krautter J., Schmitt J.H.M.M., et al. 
1996, \aap, 312, 439
\bibitem[\protect\citeauthoryear{Wichmann et al.}{2000}]{wichmann00}
Wichmann R., Torres G., Melo C.H.F., et al. 
2000, \aap, 359, 181
\bibitem[\protect\citeauthoryear{Wo\'{z}niak et al.}{2004}]{woz04}
Wo\'{z}niak P.R., Vestrand W.T., Akerlof C.W., et al.
2004, \aj, 127, 2436
\bibitem[\protect\citeauthoryear{Zucker \& Mazeh}{1994}]{zucker94}
Zucker, S., \& Mazeh, T. 1994, ApJ, 420, 806
\end{thebibliography}
\end{document}